\documentstyle[epsfig,twocolumn,aps,amssymb,graphics]{revtex}
\begin{document}
\newcommand{\beq}{\begin{equation}}
\newcommand{\eeq}{\end{equation}}
\newcommand{\bea}{\begin{eqnarray}}
\newcommand{\eea}{\end{eqnarray}}
\def\plumin{\underline{+}}
\def\minplu{{{\stackrel{\underline{\ \ }}{+}}}}
\def\bfr{{\bf r}}
\def\bfk{{\bf k}}
\def\bfq{{\bf q}}

\draft
%\twocolumn[\hsize\textwidth\columnwidth\hsize\csname @twocolumnfalse\endcsname
%\begin{title}
%\bf
\title{
Nonlinear screening in two-dimensional electron gases }
%\end{title}
\author{E. Zaremba}
%\begin{instit}
\address{
Department of Physics,
Queen's University,
Kingston, Ontario, Canada K7L 3N6 and
Donostia International Physics Center, Paseo Manuel de Lardizabal, no.
4, 20018 San Sebatian, Spain
}
%\end{instit}
\author{I. Nagy}
\address{
Department of Theoretical Physics, Institute of Physics, 
Technical University of Budapest, H-1521 Budapest, Hungary, and
Donostia International Physics Center, Paseo Manuel de Lardizabal, no.
4, 20018 San Sebatian, Spain}
\author{P. M. Echenique}
\address{
Departamento de Fisica de Materiales
and Centro Mixto Fisica Materiales CSIC/UPV,
Facultad de Quimica, Universidad del Pais Vasco, Apdo 1072,
San Sebastian, Spain
}

\date{\today}

\maketitle

\begin{abstract}
We have performed self-consistent calculations of the nonlinear
screening of a point charge $Z$ in a two-dimensional electron gas using
a density functional theory method. We find that
the screened potential for a $Z=1$ charge supports a bound state 
even in the high density limit where one might expect perturbation
theory to apply. To explain this behaviour, we prove a theorem
to show that the results of linear response theory are in fact correct
even though bound states exist.
\end{abstract}


Screening is a 
fundamental property of an electron gas in arbitrary dimensions. 
The example of two dimensions is of particular interest
because of the possible realization of quasi-two-dimensional
systems in a variety of contexts:
semiconductor heterostructures\cite{ando82}, 
image or band-gap surface states at metal surfaces\cite{echenique99}, 
electrons on the surface of liquid helium\cite{andrei97},
and layered materials\cite{layered}.
In all of these cases, the interaction of external charges with the
two-dimensional electron gas (2DEG) is a problem of both fundamental and
practical interest. For example, the transport of electrons in a 2DEG
is often limited by charged impurity scattering and a detailed knowledge
of the scattering potential is needed for an accurate determination of
the electron mobility\cite{ando82}. 
Scanning tunnelling microscopy offers an even more direct means of
determining the screening response of a quasi-2DEG through the
observation of adsorbate-induced Friedel oscillations\cite{friedel}.
Still another class of problems involves the
interaction with {\it moving} charges as might arise in low-energy
electron scattering\cite{nagao01} or tunnelling 
experiments\cite{murphy95}.  In this case, the dynamic
response of the 2DEG is important in that electronic excitation, and
hence energy loss, will occur.

A charged impurity or projectile typically represents a strong
perturbation and a nonlinear screening theory is in general needed to
account for the modifications of the local electronic structure.
However, in certain situations the {\it screened} impurity
potential may be relatively weak and therefore amenable to a
perturbative treatment. This is usually the method adopted to deal with
donor impurities that are spatially removed from the 2DEG within a
heterostructure\cite{ando82}, although it is rare to find quantitative 
agreement between theory and experimentally measured 
mobilities\cite{fletcher90}. 
The situation of
acceptor impurities {\it within} the gas is a much more severe
perturbation and quite dramatic effects can arise as a result of the
modified electronic structure\cite{richter89,zaremba91}. 
In such situations, the screening
response has to be determined nonlinearly.

One of our objectives in this Letter is to provide a fully
self-consistent description of the nonlinear screening in an 
ideal 2DEG within the context of density functional theory. A second
objective is to use these calculations to establish the range of 
validity of linear response theory. Somewhat surprisingly, this latter
objective is more subtle than anticipated as a result of a
peculiarity of potential scattering in 2D, namely the fact
that any purely attractive potential always has at least one bound
state\cite{simon76}, in marked contrast to the situation in 3D.
For a positively charged impurity we in fact encounter a situation in
which the screened potential supports a bound state even in the high
density limit where one would intuitively expect a perturbative
treatment to be valid. To resolve this apparent paradox, we prove what
will be referred to as the high density screening theorem which states
that the screening charge can indeed be calculated perturbatively 
{\it even when bound states exist}. 
In this way we are able to justify the use of
perturbation theory when at first sight it would seem inapplicable.

%As one possible application of our results we consider the low 
%velocity stopping
%power for a heavy projectile moving parallel to the plane of the 2DEG.
%This problem has been treated previously in both linear and nonlinear
%response approximations but none of these calculations is truly
%self-consistent in the sense considered here. To calculate the stopping
%power we make use of a kinetic theory formulation of the scattering of
%electrons from the self-consistent screened potential.

The problem we address is the nonlinear screening of a stationary point
charge, $Z$, located in the plane of a 2DEG. The latter
is treated as ideal in the sense that the electrons are confined to the
plane. Of course in real applications, such as a heterostructure,
a more accurate treatment of the electronic states is required.
We ignore these complications in order to focus on those aspects of 
nonlinear screening which are expected to be independent of these 
details. To stabilize
the system, the electrons move in the presence of a uniform
neutralizing positive background. In addition, we assume the electrons 
to have an isotropic effective mass $m^*$ and to be immersed 
in an extended dielectric with permittivity $\varepsilon$. We use
the effective Bohr radius, $a_0 = \varepsilon \hbar^2/m^*e^2$, as the
unit of length and the effective Hartree, $H = e^2/\varepsilon a_0$, as
the unit of energy. The density
of the gas, $n_0$, is characterized by the density parameter
$r_s = 1/\sqrt{\pi n_0}$.

The static screening response of the 2DEG is determined by solving
self-consistently the two-dimensional Kohn-Sham equations
\beq
-{1\over 2} \nabla^2 \psi_i(\bfr) + \Delta v_{\rm eff}(\bfr) 
\psi_i(\bfr)= E_i \psi_i(\bfr) 
\label{SE}
\eeq
where the effective potential is given by
\beq
\Delta v_{\rm eff}(\bfr) = v_{\rm ext}(\bfr) + \Delta v_H(\bfr)+
\Delta v_{xc}(\bfr)\,.
\label{veff}
\eeq
Here, $v_{\rm ext}=-Z/r$ is the
external potential and $\Delta v_H$ is the Hartree potential 
%\beq
%\Delta v_H(\bfr) = \int d^2r' {\Delta n(\bfr') \over |\bfr-\bfr'|}
%\label{Hartree}
%\eeq
due to the electronic screening density, $\Delta n(\bfr) = n(\bfr)-n_0$.
The change in the exchange-correlation potential, $\Delta v_{xc}(\bfr) =
v_{xc}[n(\bfr)] - v_{xc}[n_0]$, is 
defined in the local density approximation (LDA)
using the parameterization of the 2D exchange-correlation energy
given in Ref.~\cite{tanatar89}. 

The total screening charge is given by
\beq
\Delta n(\bfr) = \sum_b |\psi_b(\bfr)|^2 + \sum_i \left 
[|\psi_i(\bfr)|^2 - |\psi_i^0(\bfr)|^2 \right ]
\eeq
where the first sum extends over all bound states of the effective
potential, and the second extends over all occupied continuum states up
to the Fermi level $E_F$. We assume that each spatial orbital is
doubly-occupied for spin. The scattering states $\psi_i(\bfr)$ are
shifted asymptotically relative to the free-particle solutions
$\psi_i^0(\bfr)$ by a phase $\eta_m(E)$. These
scattering phase shifts are related to the total screening charge 
according to the 2D Friedel sum rule (FSR)\cite{stern67}
\beq
 Z_{FSR} = {2\over \pi}\sum_{m=-\infty}^\infty \eta_m(E_F)\,.
\label{FSR}
\eeq
Details of the self-consistent solution of Eqs.~(\ref{SE}) and 
(\ref{veff}) will be presented elsewhere.
%obtained using standard numerical techniques~\cite{zaremba02}. 

We begin by considering the case of a negatively charged impurity
($Z=-1$), such as an antiproton or acceptor state in a semiconductor.
Fig.~1 presents the self-consistent effective potential 
$\Delta v_{\rm eff}$ for $r_s=4$ (solid curve) as a function of the
distance from the impurity. This potential repels electrons almost 
completely from the impurity's vicinity, leaving
exposed a positively charged disc of radius $R \simeq r_s$
which neutralizes the impurity charge. As is the case in 3D, this 
behaviour cannot be reproduced in a linear theory.
The screened $Z = -1$ potential does not support bound states for any 
$r_s$ value, as might have been expected. We mention this since it 
has been claimed\cite{ghazali95} 
that the introduction of a negative test charge into a 2D
gas can give rise to a potential which is sufficiently strong to bind 
an electron. 

\begin{figure}
\centering
\psfig{file=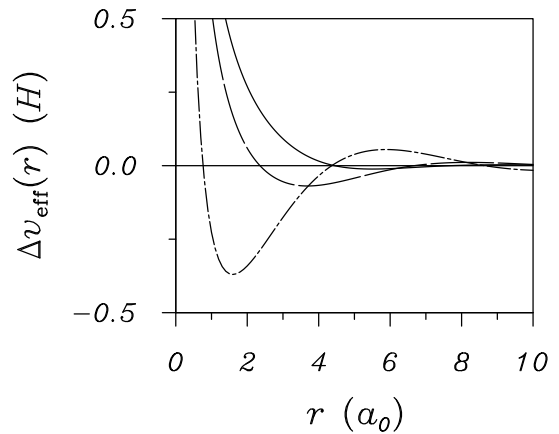, scale=0.40, bbllx=30, bblly=150,
 bburx=540, bbury=570} 
\caption{\label{potentials}
 Comparison of nonlinear and linear screened potentials: nonlinear with
exchange-correlation (solid), nonlinear Hartree potential 
(short-dashed), linear Hartree potential with LFC (chain); $Z=-1$, 
$r_s =4$.}
\end{figure}

To make contact with this
earlier work we compare in Fig.~\ref{potentials}
the nonlinearly screened potentials
with those obtained on the basis of linear response theory. The chain
curve shows the linear response Hartree potential ($v_{\rm ext} + 
\Delta v_H$) as obtained when local 
%shows the screened potential as obtained when only Hartree interactions
%are included (the random phase approximation).
%A second curve shows the Hartree potential as obtained when local 
field corrections (LFC) are included in the determination of the
electron screening density\cite{ghazali95}. 
This potential has a large attractive
region in real space and supports a bound state for a unit negative
test charge of one electron mass. This observation led to the
suggestion of a possible pairing mechanism that could be responsible 
for a correlation-induced instability at low densities\cite{ghazali95}.
Such a conclusion, however, is
invalid on two counts. First, the screening of the impurity is 
{\it strongly nonlinear}. The dashed curve in Fig.~\ref{potentials} 
shows the corresponding Hartree potential when the nonlinear
screening density is used. It has a much shallower attractive region.
But more importantly, an electron,
as opposed to a negative test charge, also feels the effect of the 
induced xc potential. With this contribution included in the full
nonlinear potential $\Delta v_{\rm eff}$ (solid curve),
there is no tendency for bound state formation, as 
confirmed numerically.

The results for a positive impurity are quite different in that the
attractive screened potential supports bound states for all densities of
physical interest (including $r_s \to 0$). 
For $Z=1$ there is one $m=0$ bound state that is doubly
occupied. Since the total screening density integrates to unity to
satisfy the FSR, the continuum screening density must itself contribute
a total charge of +1 in order to compensate for the overscreening
provided by the bound states. In other words, the $Z=1$ impurity with 2
bound electrons can be viewed as an H$^-$ ion which acts as a $Z=-1$
impurity. This was confirmed by comparing the $Z=1$ and $Z=-1$ {\it
continuum} screening densities. For $r_s = 10$, the two are
virtually the same.

\begin{figure}
\centering
\psfig{file=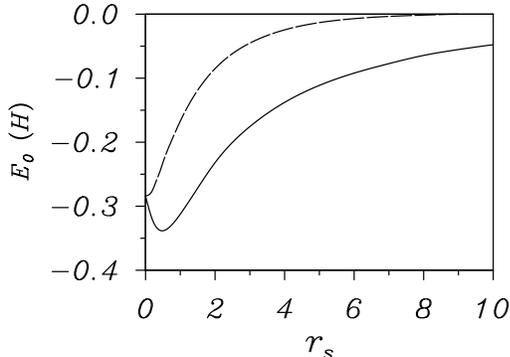, scale=0.38, bbllx=60, bblly=200,
 bburx=575, bbury=575} 
\caption{\label{eigenvalue}
 Bound state eigenvalue {\it vs.} $r_s$: nonlinear DFT with (solid) and
without (dashed) exchange-correlation.}
\end{figure}

In Fig.~\ref{eigenvalue} we show the $Z=1$ bound state energy as a 
function of $r_s$. 
The behaviour seen is surprising in view of the corresponding behaviour
in  3D. There, the
bound state energy {\it increases} with decreasing $r_s$ since the
impurity potential is screened more effectively with increasing density
and, as a result, the bound state eventually ceases to
exist\cite{zaremba77}.
Beyond this point, the accuracy of perturbation theory improves with
increasing density. The contrary behaviour exhibited in 
Fig.~\ref{eigenvalue} calls into question the applicability of 
perturbation theory in the 2D case. 

To address this question,
we now prove a theorem
regarding electronic screening in the high density limit. We consider
the introduction of an external potential $\lambda V(\bfr)$ into a
uniform {\it noninteracting} Fermi gas for arbitrary dimension $D$. The
parameter $\lambda$ is a coupling constant whose physical
value is unity. The problem is to determine the induced density $\Delta
n(\bfr)$ due to the introduction of $\lambda V(\bfr)$; the effect of
interactions will be dealt with afterwards.
By making use of Dyson's equation, $G_{\lambda+\delta \lambda} =
G_\lambda + \delta \lambda G_\lambda V G_{\lambda+\delta \lambda}$,
for the single particle Green
function $G_\lambda(\bfr,\bfr',z)$, one can show that the screening
density satisfies
\bea
{\partial \Delta n(\bfr; \lambda) \over \partial \lambda} =
&-&{1\over \pi} \int d\bfr' V(\bfr') 
\int_{-\infty}^{E_F} dE\,\nonumber \\
&\times& \mbox {Im} \left [G_\lambda(\bfr,\bfr',E+i\epsilon)
G_\lambda(\bfr',\bfr,E+i\epsilon)\right ]\,.
\eea
The crucial next step in the argument is to
use the analytic properties of the Green functions to change
the energy integral from $(\int_{-\infty}^{E_F})$ to 
$(-\int_{E_F}^\infty)$. In the $E_F \to \infty$ limit
it is then permissible to replace $G_\lambda$ by the free-particle
Green function $G_0$ since the energy
$E$ can now be assumed to be much larger than the strength of the
(bounded) potential $|V(\bfr)|$. Once this is done, the energy
integration can be changed back to its original range, and after
integrating with respect to the coupling constant, we obtain
\bea
\Delta n(\bfr) \simeq 
&-&{1\over \pi} \int d\bfr' V(\bfr') 
\lim_{E_F \to \infty} \int_{-\infty}^{E_F} dE\, \nonumber \\
&\times&\mbox{Im} \left [
G_0(\bfr,\bfr',E+i\epsilon) G_0(\bfr',\bfr,E+i\epsilon)\right ]\,.
\label{asympt}
\eea
This asymptotic result is valid for any $D$ and applies even 
when the potential $V(\bfr)$ supports bound states.
 
An alternative expression for (6) is
\beq
\Delta n(\bfr) \simeq  -\lim_{E_F \to \infty}{1\over L^D} \sum_{\bfq}
\chi_0(\bfq) V(\bfq) e^{i\bfq \cdot \bfr}
\label{17}
\eeq
which is the result of linear response theory. Here,
$\chi_0(\bfq)$ is the noninteracting static density response
function of the system. In the high density limit, $q \ll k_F$
for all wavevectors for which $V(\bfq)$ is finite, and we have $\Delta
n(\bfr) \simeq - \chi_0(0) V(\bfr)$. In 2D, $\chi_0(0) = \pi^{-1}$
and
\beq
\Delta n(\bfr) \simeq  -{1\over \pi} V(\bfr)\,.
\label{asympt_real}
\eeq
Thus we arrive at the interesting conclusion that in 2D the screening
density takes on a {\it density-independent} form in the high density 
limit, and is simply proportional to the perturbing potential.

\begin{figure}
\centering
\psfig{file=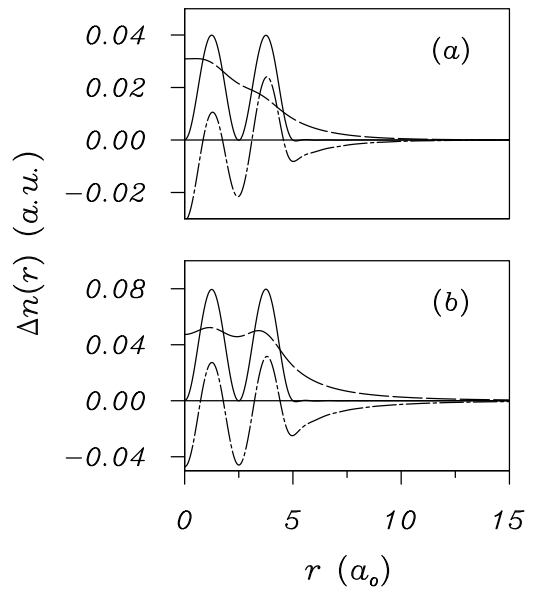, scale=0.40, bbllx=40, bblly=115,
 bburx=550, bbury=700} 
\caption{\label{model}
 Screening charge density for a model potential: bound (dashed) and
continuum (chain) state contributions, total (solid).
(a) $V_0=0.125$ H, (b) $V_0=0.25$ H; $r_s = 0.5$.}
\end{figure}

This result can easily be checked numerically. In Fig.~\ref{model} we
show $\Delta n(\bfr)$ for a 2D gas with $r_s = 0.5$ for a model
potential $V(r) = -V_0 \sin^2(2\pi r/r_0)\theta(r_0-r)$ which is an
axially symmetric, double-well potential. With $r_0 = 5$
a.u. and $V_0 = 0.125$  H the potential has a single $m=0$ bound state
while for $V_0 = 0.25$ H there are two $m=0$ bound states. In both cases
we see that the total screening density is well-approximated by the
asymptotic result in Eq.~(\ref{asympt_real}).

To include the effect of interactions, we identify $\Delta v_{\rm eff}$
in Eq.~(\ref{veff}) with $V(r)$ and make use of Eq.~(\ref{asympt_real}).
In the high-density limit we then find
\beq
\Delta v_{\rm eff}(q) = - {2\pi Z \over q+2}\,,
\eeq
which in real space gives the Thomas-Fermi potential\cite{stern67}. 
This potential is
purely attractive and has a bound state eigenvalue of $E_0 = -0.2862$ H
which is the $r_s\to 0$ limit of the curves in Fig.~\ref{eigenvalue}. 
This explains why a bound state exists in the high density limit
and why, in spite of this, the result is not in conflict with the
applicability of linear response theory. It should be emphasized that
the same argument in 3D leads to the conclusion that no bound state
can exist in this case.

As a final practical application we present results of the
calculation of the energy loss per unit length (or stopping power, $S$)
for a projectile of charge $Z$ moving with velocity $v$ in the plane
of the 2DEG.
Within the so-called kinetic theory framework, the 
stopping power is given by the expression\cite{bonig89}
\beq
S = n_0 v v_F \sigma_{tr}(E_F)\,,
\eeq
where $\sigma_{tr}(E_F)$ is
the momentum-transfer cross section defined in terms of the scattering
phase shifts by\cite{stern67}
\beq
\sigma_{tr}(E_F) = {4\over v_F} \sum_{m=0}^\infty \sin^2[\eta_m(E_F)-
\eta_{m+1}(E_F)]\,.
\eeq
To leading order in the velocity, it is sufficient to determine the
scattering phase shifts using the static nonlinearly screened potentials
calculated in the present paper.

\begin{figure}
\centering
\psfig{file=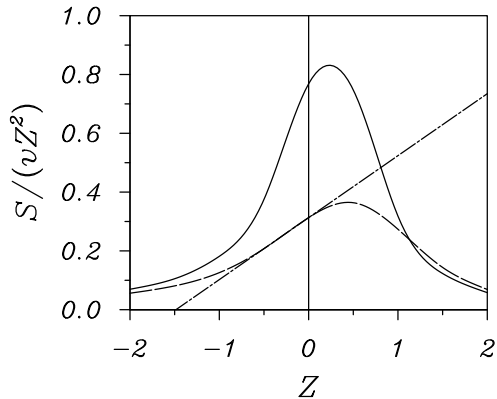, scale=0.4, bbllx=65, bblly=175,
 bburx=550, bbury=580} 
\caption{\label{stopping}
 Normalized stopping power as a function of the projectile charge $Z$,
for $r_s=2$: nonlinear screening with (solid) and without
(dashed) exchange-correlation. The straight chain curve gives the
quadratic response result.}
\end{figure}

In Fig.~\ref{stopping} we show the stopping power as a function of the 
projectile charge $Z$.
For small $Z$, $S$ has the expansion $S = S_1 Z^2 + S_2 Z^3+\dots$
where the first two terms are the linear and quadratic response results,
respectively. To emphasize the deviations from linear response, we
present the results in the form $S/(vZ^2)$. In this representation,
the stopping power including the quadratic response correction appears
as a straight line with slope $S_2/v$. This correction was previously
calculated within the quadratic  random phase
approximation\cite{bergara99} and is shown in Fig.~\ref{stopping}
as the straight line. We can see that corrections beyond quadratic
response theory are large, especially for positive charges. 
Furthermore, the
inclusion of xc is seen to enhance the stopping power considerably in 
the range $-1 \le Z \le 1$, even to a greater extent than found in 
3D\cite{echenique91}. Finally, comparison with earlier
calculations\cite{bret93} 
demonstrates that important differences arise when the self-consistently
determined nonlinear screening potentials are used to evaluate the
momentum scattering cross-sections.

In summary, we have performed self-consistent calculations of 
the nonlinear screening 
of a point charge in a 2DEG using density functional theory. We have
also proved a screening theorem which clarifies the behaviour of the
screening in the high density limit. These results find application in a
variety of problems, such as charged impurity scattering and the
stopping power of charged projectiles.

%\acknowledgments
The work of I.N. has been supported by the OTKA (Grant Nos.
T025019 and T029813), and that of E.Z. by a grant from the Natural 
Sciences and Engineering Research Council of Canada. P.M.E. thanks the
University of the Basque Country, the Basque Hezkuntza, Unibertsitate
eta Ikerketa Saila and the Spanish MCyT for support.


\begin{references}
\bibitem{ando82} T. Ando, A. B. Fowler and F. Stern, Rev. Mod. Phys.
{\bf 54}, 437 (1982).
\bibitem{echenique99} P. M. Echenique {\it et al.},
Chem. Phys. {\bf 251}, 1 (1999).
\bibitem{andrei97} {\it Two-Dimensional Electrons on Cryogenic
Substrates}, ed. by E. Andrei (Kluwer Academic Press, Dordrecht, 1997).
\bibitem{layered} M. S. Dresselhaus and G. Dresselhaus, Adv. Phys.
{\bf 51}, 1 (2002); J. Singleton and C. Mielke, Contemp. Phys. {\bf
43}, 63 (2002).
\bibitem{friedel} M. F. Crommie {\it et al.}, Nature {\bf
363}, 524 (1993); P. A. Avouris {\it et al.}, J. Vac. Sci. Tech. B{\bf
12}, 1447 (1994); P. T. Sprunger {\it et al.}, Science {\bf 275}, 1764
(1997).
\bibitem{nagao01} T. Nagao {\it et al.},
Phys. Rev. Lett. {\bf 86}, 5747 (2001).
%\bibitem{nagao01} T. Nagao, T. Kildebrandt, M. Henzler and S. Hasegawa,
%Phys. Rev. Lett. {\bf 86}, 5747 (2001).
\bibitem{murphy95} S. Q. Murphy {\it et al.}, Phys. Rev. B {\bf 52},
14825 (1995).
\bibitem{fletcher90} R. Fletcher {\it et al.},
Phys. Rev. B {\bf 41}, 10649 (1990).
\bibitem{richter89} J. Richter {\it et al.},
Phys. Rev. B {\bf 39}, 6268 (1989). 
\bibitem{zaremba91} E. Zaremba, Phys. Rev. B {\bf 44}, 1379 (1991).
\bibitem{simon76} B. Simon, Ann. Phys. (N.Y.) {\bf 97}, 279 (1976).
\bibitem{tanatar89} B. Tanatar and D. M. Ceperley, Phys. Rev. B 
{\bf 39}, 5005 (1989).
\bibitem{stern67} F. Stern and W. E. Howard, Phys. Rev. {\bf 163},
816 (1967).
%\bibitem{zaremba02} E. Zaremba, I. Nagy and P. M. Echenique, to be
%published.
\bibitem{ghazali95} A. Ghazali and A. Gold, Phys. Rev. B {\bf 52},
16634 (1995).
\bibitem{zaremba77} E. Zaremba {\it et al.},
J. Phys. F {\bf 7}, 1763 (1977).
\bibitem{bonig89} L. B\"onig and K. Sch\"onhammer, Phys. Rev. B {\bf
39}, 7413 (1989).
\bibitem{bergara99} A. Bergara {\it et al.}, Phys. 
Rev. B {\bf 59}, 10145 (1999).
\bibitem{echenique91} P.M. Echenique {\it et al.},
Nucl. Instrum. Methods Phys. Res. B {\bf 56/57}, 345 (1991).
%\bibitem{vinter82} B. Vinter, Phys. Rev. B {\bf 26}, 6808 (1982).
%\bibitem{fetter73} A. Fetter, Ann. Phys. (N.Y.) {\bf 81}, 367 (1973).
%\bibitem{horing87} N.J.M. Horing, H.C. Tso, and G. Gumbs, Phys. Rev. 
%B {\bf 36}, 1588 (1987).
%\bibitem{wang95} Y.-N. Wang and T.-C. Ma, Phys. Rev. B {\bf 52}, 
%16395 (1995).
%\bibitem{bergara97} A. Bergara, I. Nagy, and P.M. Echenique, Phys. Rev.
%B {\bf 55}, 12864 (1997).
\bibitem{bret93} A. Bret and C. Deutsch,  Phys. Rev. E {\bf 48}, 2994
(1993); A. Krakovsky and J. K. Percus, Phys. Rev. B {\bf 52}, R2305 
(1995); Y.-N. Wang and T.-C. Ma, Phys. Rev. A {\bf 55}, 2087 (1997).
%\bibitem{hu88} C.D. Hu and E. Zaremba, Phys. Rev. B {\bf 37}, 9268
%(1988).
%\bibitem{nagy95} I. Nagy, Phys. Rev. B {\bf 51}, 77 (1995).
%\bibitem{wang97} Y.-N. Wang and T.-C. Ma, Phys. Rev. B {\bf 55}, 
%2087 (1997).
%\bibitem{sjolander72} A. Sj\"olander and M.J. Stott, Phys.
%Rev. B {\bf 5}, 2109 (1972).
%\bibitem{krakovsky95} A. Krakovsky and J. K. Percus, Phys. Rev. B {\bf
%52}, R2305 (1995).
%\bibitem{echenique96} P.M. Echenique and E. Zaremba, Physicalia Mag.
%{\bf 18}, 229 (1996).
%\bibitem{nagy89} I. Nagy, A. Arnau, P.M. Echenique, and E. Zaremba, 
%Phys. Rev. B {\bf 40}, 11983 (1989).
%\bibitem{adhikari86} S. K. Adhikari, Am. J. Phys. {\bf 54}, 362
%(1986).
%\bibitem{shore77} H.B. Shore, J.H. Rose, and E. Zaremba, Phys. Rev. B
%{\bf 15}, 2858 (1977).
%\bibitem{footnote1} We remark that the sign in Eq.~(\ref{epsRPA}) is
%often seen to be negative as a result of a different choice of sign for
%the response function $\chi_0(q)$.
%\bibitem{echenique91} P.M. Echenique, A. Arnau, M. Pe\~nalba, and I.
%Nagy, Nucl. Instrum. Methods Phys. Res. B {\bf 56/57}, 345 (1991).
%\bibitem{krakovsky95b} A. Krakovsky and J. K. Percus, Phys. Rev. B {\bf
%52}, 7901 (1995).
%\bibitem{butler62} D. Butler, Proc. Phys. Soc. {\bf 80}, 741 (1962).
%\bibitem{kohn65} W. Kohn and C. Majumdar, Phys. Rev. {\bf 138}, A1617
%(1965).
%\bibitem{S67} F. Stern, Phys. Rev. Lett. {\bf 18}, 546 (1967).
%\bibitem{kliewer00} J. Kliewer, R. Berndt, E. V. Chulkov, V. M. Silkin,
%P. M. Echenique and S. Crampin, Science {\bf 288}, 1399 (2000).
%\bibitem{huefner96} S. H\"ufner, {\it Photoelectron Spectroscopy},
%(Springer, Berlin, 1996), 2nd ed.

\end{references}
\end{document}